\documentclass[prl,twocolumn,amsmath,showpacs,superscriptaddress]{revtex4}
\usepackage{bm}
\usepackage{graphicx}

\newcommand{\dpar}[2]{\frac{\partial #1}{\partial #2}}
\newcommand{\eps}{\varepsilon}
\newcommand{\bracket}[1]{\langle #1 \rangle}
\newcommand{\ket}[1]{\lvert #1 \rangle}

\begin{document}

\title{Berry phase effect in anomalous thermoelectric transport}

\author{Di Xiao}
\affiliation{Department of Physics, The University of Texas, Austin,
  Texas 78712, USA}
\author{Yugui Yao}
\affiliation{Beijing National Laboratory for Condensed Matter Physics, 
  Institute of Physics, Chinese Academy of Sciences,
  Beijing, 100080, China}
\author{Zhong Fang}
\affiliation{Beijing National Laboratory for Condensed Matter Physics,
  Institute of Physics, Chinese Academy of Sciences,
  Beijing, 100080, China}
\author{Qian Niu}
\affiliation{Department of Physics, The University of Texas, Austin,
  Texas 78712, USA}

\date{May 29, 2006}

\begin{abstract}

We develop a theory of Berry phase effect in anomalous transport in
ferromagnets driven by statistical forces such as the gradient of
temperature or chemical potential.  Here a charge Hall current arises
from the Berry phase correction to the orbital magnetization rather
than from the anomalous velocity which does not exist in the absence
of a mechanical force.  A finite-temperature formula for the orbital
magnetization is derived, which enables us to provide an explicit
expression for the off-diagonal thermoelectric conductivity, to
establish the Mott relation between the anomalous Nernst and Hall
effects, and to reaffirm the Onsager relations between reciprocal
thermoelectric conductivities.  A first-principles evaluation of our
expression is carried out for the material CuCr$_2$Se$_{4-x}$Br$_x$,
obtaining quantitative agreement with a recent experiment.
\end{abstract}

\maketitle

The phenomena of transport fall into two categories: those due to a
mechanical force, such as the electric field on charges, and those
driven by a statistical force, such as the gradient of temperature or
chemical potential.  The mechanical force exists on the microscopic
level and can be described by a perturbation to the Hamiltonian for
the carriers, while the statistical force manifests on the macroscopic
level and makes sense only through the statistical distribution of the
carriers.  It has been established~\cite{chang1996,sundaram1999} that
Berry phase of Bloch states has a profound effect on transport driven
by a mechanical force.  This is through the mechanism that the group
velocity of a Bloch electron acquires an anomalous term proportional
to the mechanical force, i.e.,
\begin{equation} \label{EOM}
\dot{\bm r} = \frac{1}{\hbar}\dpar{\eps_n(\bm k)}{\bm k}
+ \frac{e}{\hbar}\bm E\times\bm\Omega_n(\bm k) \;,
\end{equation}
where $\eps_n(\bm k)$ is the band energy, $-e\bm E$ is the mechanical
force due to the external electric field, and $\bm\Omega_n(\bm k)$ is
the Berry curvature, the Berry phase per unit area in the $\bm
k$-space.  Evaluation of the Hall current from the anomalous term
reproduces the Karplus-Luttinger formula~\cite{karplus1954} for the
anomalous Hall conductivity.  Calculations based on the Berry phase
effect have found much success in explaining anomalous Hall effects
(AHE) in ferromagnets of semiconductors~\cite{jungwirth2002},
oxides~\cite{fang2003} and transition metals~\cite{yao2004}.  Recent
experiments~\cite{lee2004,zeng2006} give further convincing evidence
in support of this theory.

A natural question is whether and how the Berry phase also manifests
in transport driven by a statistical force.  On the one hand, the
anomalous velocity vanishes in the absence of a mechanical force,
eliminating the obvious cause for a Berry phase effect in this case.
On the other hand, this conclusion would introduce a number of basic
contradictions to the standard transport theory.  First, a chemical
potential gradient would be distinct from the electrical force,
violating the basis for the Einstein relation for transport.  Second,
a temperature gradient would not induce an intrinsic charge Hall
current, violating the Mott relation [see Eq.~\eqref{mott} below]
between the AHE and the anomalous Nernst effect (ANE), where a
transverse current is produced by a temperature gradient in
ferromagnets.  Third, as will be made clear below, it would be
impossible to establish the Onsager relation between cross transport
coefficients connecting thermoelectric Hall currents and forces.  In
addition, a recent experiment on the ANE in the spinel ferromagnet
CuCr$_2$Se$_{4-x}$Br$_x$~\cite{lee2004a} found weak dependence on
scattering, suggesting that there should indeed be a Berry-phase
induced intrinsic mechanism.

In this Letter, we solve the puzzle by showing how the Berry phase
effect manifests in thermoelectric transport driven by a statistical
force.  It turns out that the local current of carriers acquires an
extra term from the carrier magnetic moment in the presence of a
non-uniform distribution which arises from the gradient of temperature
or chemical potential.  However, the complete theory also relies on a
proper deduction of magnetization current~\cite{cooper1997}, and
requires a deeper understanding of the orbital magnetization. It was
found that there is a Berry-phase correction to the magnetization
~\cite{xiao2005,thonhauser2005}, and here we generalize it to the case
of finite temperatures which is needed for thermoelectric transport.
This Berry phase correction eventually enters into the transport
current produced by the statistical force, playing the counterpart as
the anomalous velocity term due to a mechanical force.

We have thus found perfect harmony between statistical and mechanical
forces even in the presence of Berry phase effect.  The basic
transport relations of Einstein, Mott, and Onsager continue to hold,
which gives strong support for the validity of our theory.  Finally,
we also provide a reality check on the Berry phase effect in the ANE
by calculating the intrinsic anomalous Nernst
conductivity $\alpha_{xy}$~\cite{alpha_xy} for
CuCr$_2$Se$_{4-x}$Br$_x$ using first-principles method.  The obtained
doping dependence curve agrees well with available experimental
data~\cite{lee2004a}.  Our calculation also predicts a peak-valley
structure between the data points, at a place where the anomalous Hall
conductivity has a sudden sign and magnitude change.

\textit{Local and transport currents.}---In the conventional Boltzmann
transport theory, one considers a statistical distribution $ g(\bm r,
\bm k)$ of carriers in the phase space of position and crystal
momentum.  The distribution function satisfies the Boltzmann equation
with a collision integral whose form depends on the details of the
collision process.  The current density is given by $\bm J =
-e\int[d\bm k]\, g(\bm r, \bm k) \dot{\bm r}$, where $\int[d\bm k]$ is
a shorthand for $\int d\bm k/(2\pi)^3$, and a summation over band
index has been omitted for simple notation.  In the absence of a
mechanical force, the electron velocity is simply $\dot{\bm r} =
\partial\eps(\bm k)/\hbar\partial\bm k$.  It is then apparent that the
anomalous velocity term due to the Berry phase drops out of the
expression for the current.

\begin{figure}
\includegraphics[width=5.5cm]{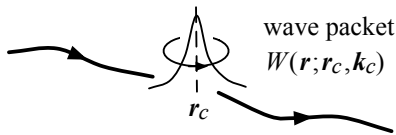}
\caption{\label{wavepacket}The wave packet description of a charge
  carrier whose center is $(\bm r_c, \bm k_c)$.  A wave packet
  generally possesses two kinds of motion: the center of mass motion
  and the self-rotation around its center.  Both of them contribute to
  the local current density as given in Eq.~\eqref{local}.}
\end{figure}

However, the above picture is na{\"i}ve that the carrier is treated as
a structureless point particle.  The quantum representation of the
carrier is in fact a wave packet, which has a finite spread in the
phase space.  The wave packet generally rotates about its center
position, as illustrated in Fig.~\ref{wavepacket}, giving rise to an
orbital magnetic moment $\bm m(\bm k) = -(e/2)\bracket{W|(\hat{\bm r}
- \bm r_c)\times\hat{\bm v}|W}$, where $\ket{W}$ is the wave packet
and $\hat{\bm v}$ is the velocity
operator~\cite{chang1996,sundaram1999}.  A careful coarse graining
analysis~\cite{culcer2004} shows that the correct expression for the
\emph{local} current has an extra term:
\begin{equation} \label{local}
\bm J = -e\int[d\bm k]\,g(\bm r, \bm k)\dot{\bm r} + \bm\nabla \times
\int [d\bm k]\,f(\bm r, \bm k) \bm m(\bm k) \;,
\end{equation}
where the magnetic moment enters explicitly.  In the extra term we
have replaced $g(\bm r, \bm k)$ with the local equilibrium Fermi-Dirac
distribution $f(\bm r, \bm k)$ for a linear-order calculation.  When
the temperature or chemical potential varies in space, the extra term
will be proportional to the gradient of these thermodynamic quantities
and is therefore non-negligible.

For transport studies, it is important to discount the contribution
from the magnetization current, a point which has attracted much
discussion in the past.  It was argued that the magnetization current
cannot be measured by conventional transport experiments (For a recent
most comprehensive work, see Ref.~\cite{cooper1997}).  Therefore, one
introduces the concept of \emph{transport} current, defined by
\begin{equation} \label{currents}
\bm j = \bm J - \bm\nabla \times \bm M(\bm r) \;,
\end{equation}
where $\bm M(\bm r)$ is the magnetization density.  This is entirely
analogous to the classic distinction between microscopic current and
free current \cite{jackson-EM}.

It is also important to realize that the magnetization density is not
simply a statistical sum of the carrier magnetic moments.  It has been
shown recently that there is a Berry phase correction to the
magnetization~\cite{xiao2005,thonhauser2005}.  The contribution from
the carrier magnetic moments to the local current will be subtracted
out in the transport current, but the Berry phase correction to the
magnetization will give rise to an extra term in the transport
current.  Earlier work concentrate on the zero temperature
magnetization, and we provide an extension to the finite temperature
case below.  Using Eq.~\eqref{magnetization} for the magnetization, we
find that the transport current is given by
\begin{equation} \label{transport}
\begin{split}
\bm j &= -e\int[d\bm k] g(\bm r, \bm k)\dot{\bm r} \\
&\quad - \bm\nabla \times \frac{1}{\beta} \int[d\bm k] \frac{e}{\hbar}
\bm\Omega(\bm k) \log(1 + e^{-\beta(\eps-\mu)}) \;,
\end{split}
\end{equation}
where $\beta = 1/k_BT$, and the Berry curvature is defined by
$\bm\Omega(\bm k) = \bm\nabla_{\bm k}\times \bracket{u|i\bm\nabla_{\bm
k}|u}$ with $\ket{u}$ being the periodic amplitude of the Bloch wave.

The above expression gives a complete account of the transport current
in ferromagnets, and for crystals with nonzero Berry curvatures in
general.  The first term is the usual expression for the charge
current, which vanishes at local equilibrium (assuming the absence of
a mechanical force), i.e., $g(\bm r, \bm k)= f(\bm r,\bm k)$.
Nonequilibrium correction to first order in the gradient of
temperature or chemical potential yields a result strongly depending
on the relaxation process, and a transverse current can result from
skew scattering due to spin-orbit coupling~\cite{smit1958}.  The
second term is new, which results from the Berry phase correction to
the magnetization.  It is also first order in the statistical force,
but is independent of the relaxation time, and is therefore an
intrinsic property of the system.

\textit{Orbital magnetization at finite temperatures.}---The orbital
magnetization of Bloch electrons has been an outstanding problem in
solid state physics.  Recently, different
approaches~\cite{xiao2005,thonhauser2005} have been used to derive a
formula at zero temperature, where Berry phase is found to play an
important role.  In order to study thermoelectric transport, we need
to generalize it to finite temperatures.  Our derivation is made easy
by using the field-dependent density of states introduced in
Ref.~\cite{xiao2005}, where it was shown that in the weak-field limit,
a quantum-state summation $\sum_{\bm k}\mathcal O(\bm k)$ of some
physical quantity $\mathcal O(\bm k)$ should be converted to a $\bm
k$-space integral according to $\int[d\bm k](1 + e\bm B \cdot \bm
\Omega/\hbar)\mathcal O(\bm k)$.

The equilibrium magnetization density can be obtained from the grand
canonical potential, which, within first order in the magnetic field,
may be written as
\begin{equation} \label{grand}
\begin{split}
F &= -\frac{1}{\beta} \sum_{\bm k} 
\log(1 + e^{-\beta(\eps_M - \mu)}) \\
&= -\frac{1}{\beta} \int[d\bm k] (1 + \frac{e}{\hbar}\bm B \cdot 
\bm\Omega) \log(1 + e^{-\beta(\eps_M - \mu)}) \;,
\end{split}
\end{equation}
where the electron energy $\eps_M = \eps(\bm k) - \bm m(\bm k) \cdot
\bm B$ includes a correction due to the orbital magnetic moment $\bm
m(\bm k)$.  The magnetization is then the field derivative at fixed
temperature and chemical potential, $\bm M = -(\partial F/\partial \bm
B)_{\mu,T}$, with the result
\begin{equation} \label{magnetization}
\begin{split}
\bm M(\bm r) &= \int[d\bm k]\, f(\bm r, \bm k) \bm m(\bm k)  \\
&\quad + \frac{1}{\beta} \int[d\bm k]\, \frac{e}{\hbar}
\bm\Omega(\bm k) \log(1 + e^{-\beta(\eps-\mu)}) \;.
\end{split}
\end{equation}
For generality, we have included a position dependence to cover the
situation of local equilibrium with a position dependent temperature
and chemical potential.

We have thus derived a general expression for the equilibrium orbital
magnetization density, valid at zero magnetic field but at arbitrary
temperatures.  The first term is just a statistical sum of the orbital
magnetic moments of the carriers originating from self-rotation of the
carrier wavepackets.  It has been derived in
Ref.~\cite{chang1996,sundaram1999} with the expression $\bm m(\bm k) =
-i(e/2\hbar) \bracket{\bm\nabla_{\bm k}u|\times[\hat{H}(\bm k) -
\eps(\bm k)]|\bm\nabla_{\bm k}u}$, where $\hat{H}(\bm k)$ is the
crystal Hamiltonian acting on $\ket{u}$.  It has the same symmetry
properties as the Berry curvature.  The second term of
Eq.~\eqref{magnetization} is the Berry phase correction to the orbital
magnetization.  It is of topological nature, arising from a bulk
consideration on the one hand as in the above derivation, and being
connected to a boundary current circulation on the
other~\cite{xiao2006b}.  Interestingly, it is this second term that
eventually enters the transport current.

\textit{Anomalous thermoelectric transport.}---With the aid of
Eq.~\eqref{transport} it is straightforward to calculate various
thermoelectric response to statistical forces.  For example, a
chemical potential gradient will produce, through the second term, a
Hall current given by $-\bm\nabla\mu\times(e/\hbar)\int[d\bm k]
f(\bm k) \bm\Omega(\bm k)$.  This is the same as the Berry-phase
induced anomalous Hall current in response to an electric field if one
substitutes $\bm\nabla\mu/e$ for the field.  It is gratifying to see
that the Einstein relation continues to hold in the presence of the
Berry phase effect.

In the presence of a temperature gradient, an intrinsic Hall current
also results from the second term of Eq.~\eqref{transport},
\begin{equation} \label{nernst} \begin{split}
\bm j_\text{in} &= -\frac{\bm\nabla T}{T} \times \frac{e}{\hbar}
\int[d\bm k] \bm \Omega [(\eps-\mu)f  \\
&\qquad+ k_BT\log(1+e^{-\beta(\eps-\mu)})] \;.
\end{split} \end{equation}
One can then extract an anomalous Nernst conductivity $\alpha_{xy}$
defined by $j_x = \alpha_{xy}(-\nabla_y T)$.  On a different route, we
can also obtain the same result by invoking a fictitious gravitational
field~\cite{luttinger1964}, establishing the Einstein relation between
this mechanical force and the temperature gradient.

Interestingly, by integration by parts, $\alpha_{xy}$ can be written
into the following more suggestive form
\begin{equation} \label{alpha}
\alpha_{xy} = -\frac{1}{e}\int d\eps\, \dpar{f}{\mu}
\sigma_{xy}(\eps) \frac{\eps-\mu}{T} \;,
\end{equation}
where $\sigma_{xy}(\eps)$ is the intrinsic anomalous Hall conductivity
at zero temperature with Fermi energy $\eps$, given by
\begin{equation}
\sigma_{xy}(\eps) = -\frac{e^2}{\hbar}\int[d\bm k]\,
\Theta(\eps-\eps_{\bm k}) \Omega_z(\bm k) \;.
\end{equation}
At low temperatures, the above relation reduces to 
\begin{equation}  \label{mott}
\alpha_{xy} = \frac{\pi^2}{3}\frac{k_B^2T}{e}\sigma'_{xy}(\eps_F) \;.
\end{equation}
Such relations between the electrical and thermoelectric
conductivities are known as Mott relations.  They were proved for
non-magnetic materials without or with a magnetic
field~\cite{marder-CM,wang2001}. Our result extends the validity of
this relation to ferromagnets and other systems with a Berry
curvature, and justifies the usage of Eq.~\eqref{mott} in
Ref.~\cite{lee2004a}.

The reciprocal of the ANE is the generation of a transverse heat
current by an electric field.  Onsager relation dictates that the
Berry phase should also affect the latter.  To show this explicitly,
we consider the energy current carried by a wave packet $
\bracket{W|(\hat{H}\hat{\dot{\bm r}}+\hat{\dot{\bm r}}\hat{H})/2|W} =
\eps\dot{\bm r} - \bm E\times\bm m(\bm k)$, where the second term is
from the field correction to the local Hamiltonian.  Assuming a
uniform temperature and chemical potential~\cite{rapid}, we obtain the
\emph{local} energy current to first order in the electric field:
\begin{equation}
\bm J^E = \int[d\bm k]\, g(\bm k)\eps\dot{\bm r}
- \bm E \times \int[d\bm k]\, f(\bm k) \bm m(\bm k) \;,
\end{equation}
where the electron velocity $\dot{\bm r}$ is given by Eq.~\eqref{EOM}.
However, the energy current also has a magnetization part from an
``energy'' magnetization~\cite{cooper1997}.  In the present case, it
is given by $-\bm E\times\bm M$, which is nothing but the
material-dependent part of the Poynting vector $\bm E \times \bm H$
describing the energy flow (with $\bm H = \bm B/\mu_0 - \bm
M$)~\cite{jackson-EM}.  Since this energy flow exists in an
equilibrium state, it does not correspond to a transport current thus
must be subtracted from $\bm J^E$ to yield the \emph{transport} energy
current $\bm j^E = \bm J^E + \bm E \times \bm M$.  Based on our
expression~\eqref{magnetization} for the magnetization density, we
finally find the Berry phase correction to the heat current (defined
by $\bm j^Q \equiv \bm j^E - \mu \bm j$):
\begin{equation} \begin{split}
\bm j_\text{in}^Q &= \bm E \times \frac{e}{\hbar}\int[d\bm k]
\bm\Omega[(\eps-\mu) f \\
&\qquad + k_BT \log(1+e^{-\beta(\eps-\mu)})] \;,
\end{split} \end{equation}
while the usual expression for the heat current is $\int[d\bm k]\,
g(\bm k) (\eps-\mu){\bm v}$, where $\bm v$ is the usual group velocity
determined by the band energy.  In this case, the Berry phase
correction comes from both the anomalous velocity and the orbital
magnetization.  Comparison with Eq.~\eqref{nernst} shows that the
Onsager relation is indeed satisfied, providing a strong evidence for
the validity of our theory.

\textit{Comparison with experiment.}---The intrinsic anomalous Nernst
conductivity $\alpha_{xy}$ only depends on the band structure and
Berry curvature, so it can be evaluated for crystals based on first
principles methods.  Here we report our result for
CuCr$_2$Se$_{4-x}$Br$_x$ and compare with the experiment
~\cite{lee2004a}.  The band structure and Berry curvature are
calculated following the procedures in Ref.~\cite{yao2004}, using the
generalized gradient approximation for the exchange-correlation
potential.  Such calculations are very extensive, and, to reduce the
work load, we assume that doping affects the Fermi energy but not the
band structure, which is justified for the present
compounds~\cite{yao2006}.

\begin{figure}
\includegraphics[width=\columnwidth]{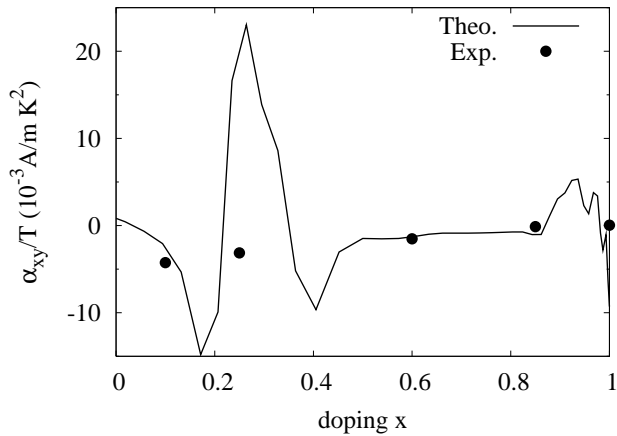}
\caption{\label{peltier}The intrinsic anomalous Nernst conductivity
$\alpha_{xy}$ (divide by the temperature $T$) of
CuCr$_2$Se$_{4-x}$Br$_x$ as a function of the Br content $x$.  The
calculated curve is compared with experimental results $\bullet$
extracted from Ref.~\cite{lee2004a}.}
\end{figure}

The calculated $\alpha_{xy}$ is plotted in Fig.~\ref{peltier} as a
function of doping $x$ together with the experimental data from
Ref.~\cite{lee2004a}.  The comparison is \emph{quantitatively} good,
except for the data point at $x=0.25$. This is however a rather
special point, because it was reported~\cite{lee2004a} that, for
unknown reasons, $\alpha_{xy}$ is not really proportional to $T$ for
$x=0.25$.  At low temperatures, a proportional relation is expected
from the Mott relation, which is followed strictly by all the data
points at other doping densities.

We also note that while our theory predicts a pronounced peak-valley
structure around $x=0.3$, the available experimental data at present
is too sparse to confirm or disprove it.  The oscillatory behavior
results from the complicated band structure of this material, and
occurs when the Fermi energy (which depends on doping) goes through a
region of spin-orbit induced energy gap.  Detailed explanation based
on the numerical calculations will be presented
elsewhere~\cite{yao2006}.  An indirect experimental evidence for this
peak is that it occurs at a place where the anomalous Hall
conductivity has a sudden change of sign and magnitude around $x=0.3$
according to Ref.~\cite{lee2004}.  Such a correlation is expected from
the Mott relation~\eqref{mott} and the fact that the Fermi energy
changes approximately linearly with the doping density~\cite{yao2006}.
Nevertheless, more direct experimental results are clearly needed for
a careful comparison with our theory.

We acknowledge useful discussions with Dimitrie Culcer, Junren Shi,
and Weida Wu.  We are grateful to Wei-Li Lee for sharing the original
experimental data and for discussions on the AHE and the ANE in
general.  D.X.\ was supported by the NSF (No.~DMR-0404252 and
DMR-0306239), Q.N.\ was supported by the DOE (No.~DE-FG03-02ER45958).
Y.G.Y.\ and Z.F.\ were supported by the NSF of China (No.~10404035,
10534030 for Y.G.Y., No.~90303022, 60576058, 10425418 for Z.F.), and
the Knowledge Innovation Project of the Chinese Academy of Sciences.

\bibliography{journal,book,footnote}

\end{document}